\documentclass[fleqn,twoside]{article}
\usepackage{espcrc2}

\usepackage{graphicx}
\usepackage[figuresright]{rotating}

\newcommand{\AmS}{{\protect\the\textfont2
  A\kern-.1667em\lower.5ex\hbox{M}\kern-.125emS}}

\title{Transverse Momentum Distribution in $b\rightarrow s\gamma$}

\author{R. Sghedoni\address[uni]{Dipartimento di Fisica, Universit\'a di Parma,\\
        Viale delle Scienze, 43100 Parma, Italia}
        \address[infn]{Gruppo Collegato INFN, Parma, Italia}}

\begin{document}

\begin{abstract}
We present the complete calculation of the transverse momentum
distribution for the decay $b\rightarrow s\gamma$. The
contributions of the leading operator $\hat{\mathcal{O}}_7$ are
computed: infrared logarithms are resummed with next-to-leading
accuracy according to usual techniques of resummation. Non
logarithmic terms are evaluated to $O(\alpha_S)$ by calculating
one loop diagrams.

\vspace{1pc}
\end{abstract}

\maketitle

\section{Introduction}
The rare decay $b\rightarrow s\gamma$ has been widely studied in
the last years, both experimentally and theoretically, for several
reasons: being loop mediated even at the lowest order, it was used
to search signals of new physics. Moreover it is a way to measure
CKM matrix elements, which is one of the main goals of $beauty$
physics to constraint the Standard Model.
\\
Perturbative QCD has been applied to $b$ physics since a long time,
because the QCD coupling constant is assumed to be small at $b$
quark mass
\begin{equation}
\alpha_S(m^2_b)\sim 0.21
\end{equation}
However a direct check of perturbative QCD is difficult to perform, because several poorly known parameters
are involved in theoretical predictions (for example CKM matrix elements and the $beauty$ quark mass). On the contrary
one usually prefers to assume that perturbative QCD is reliable in $b$ physics and can be used to extract these parameters.
\\
In our opinion \cite{noi2} it is not clear whether the
perturbative approach is viable at $b$ mass energies, both in
inclusive quantities and, \it a fortiori\rm, in semi-inclusive
distributions.
\\
In the latter case a powerful tool is the resummation of the large
logarithms appearing at the border of phase space. This approach
had a wide application in past years in processes such as
$e^+e^-$, $p\overline{p}$, deep inelastic scattering, \dots , for
semi-inclusive distributions and in recent years has become an
ordinary technique in $b$ physics too.
\\
It is well known that the resummation of large logarithms for a
generic distribution $D(x)$ gives the general result \cite{cttw}:
\begin{equation}\label{master}
D(x) = K(\alpha_S) \Sigma(x;\alpha_S) + R(x)
\end{equation}
where
\begin{itemize}
\item
$\Sigma(x;\alpha_S)$ is a universal, process independent function
resumming the infrared logarithms in exponentiated form.  In
particular
\begin{equation}\label{g}
\log \Sigma(x;\alpha_S)= L g_1(\alpha_S L) + g_2 (\alpha_S L)+
\dots
\end{equation}
being $L$ a large logarithm. The functions $g_i$ have an expansion
as
\begin{equation}
g_i(z)=\sum_1^\infty g_{i,n}z^n
\end{equation}
and resum logarithms of the same size: in particular $g_1$ resums leading logs
$\alpha_S^n \log^{n+1}$ and $g_2$ the next-to-leading ones
$\alpha_S^n \log^n$;
\item
$K(\alpha_S)$ is the coefficient function, a process dependent
function which can be calculated in perturbation theory by
evaluating the constant terms at the required order
\begin{equation}
K(\alpha_S)=1+\frac{C_F\alpha_S}{\pi} k_1 + \dots
\end{equation}
\item
$R(x)$, the remainder function, is process dependent and satisfies
the condition
\begin{equation}\label{remainder}
R(x) \rightarrow 0 \ \ \ for \ \ x \rightarrow 0
\end{equation}
It takes into account hard contributions and can be calculated in
perturbation theory as an $\alpha_S$ expansion.
\begin{equation}
R(x)=\frac{\alpha_S C_F}{\pi} r_1(x)+\dots
\end{equation}
\end{itemize}
For the previous reasons, a precise calculation of the terms
involved in this approach is needed to compare theoretical
predictions and experimental data.
\\
For $b\rightarrow s\gamma$ the resummation has been applied in the
photon spectrum distribution \cite{longitudinal}, which is
sensitive to longitudinal degrees of freedom. We have extended the
analysis to the transverse momentum distribution of the strange
quark with respect to the photon direction, which is sensitive to
transverse degrees of freedom \cite{noi}.
\\
The resummation of large logarithms seems to be required in the
transverse momentum distribution for $b\rightarrow s\gamma$,
because the double logarithmic correction becomes large near the
border of phase space, compared to 1 (lowest order contribution to
the rate):
\begin{equation}
-\frac{1}{4}\frac{C_F\alpha_S(m_b^2)}{\pi} \ \log^2{\Lambda^2\over
m^2_b}\sim\ -0.7
\end{equation}
if we push the transverse momentum to its phy\-si\-cal limit,
$\Lambda\sim 300 {\rm MeV}$, before the appearance of Landau pole
effects.
\\
The single logarithm becomes large too, due to its large numerical
coefficient
\begin{equation}
-\frac{5}{4}\frac{C_F\alpha_S(m_b^2)}{\pi} \ \log{\Lambda^2\over
m^2_b}\sim\ 0.6
\end{equation}
However hard contributions as the function $R(x)$ should be
carefully evaluated, being at these energies more relevant than in
processes such as $e^+e^-$, deep inelastic scattering, due to the
size of $\alpha_S$ \cite{noi2}.

\section{Transverse Momentum Distribution for $b\rightarrow
s\gamma$}

Let us consider the decay $b\rightarrow s\gamma$ in the $b$ rest
frame and define the photon direction as the $z-axis$. Let us
consider the transverse momentum $\vec{p}_\perp$ of the strange
quark with respect to the photon direction and, in particular, the
partially integrated distribution
\begin{equation}
D(x)=\int_0^x \ {1\over\Gamma_0} \ {d\Gamma\over dx^\prime} \
dx^\prime
\end{equation}
where $x$ is an adimensional variable defined as
\begin{equation}
x={p^2_\perp \over m^2_b}
\end{equation}
being $m_b$, the $b$ quark mass, the hard scale of the process.
$\Gamma_0$ is the Born amplitude, as defined in \cite{born}.
\\
According to (\ref{master}), an accurate calculation of $D(x)$
requires two steps:
\begin{itemize}
\item
the resummation of large logarithms arising at the border of phase
space (\it infrared region\rm) \cite{noi};
\item
a fixed order calculation to extract non lo\-ga\-rithmic
contributions, such as constants, contained in $K(\alpha_S)$, and
functions vanishing for $x\rightarrow 0$, contained in $R(x)$
\cite{noi2}.
\end{itemize}

\subsection{Resummation of infrared logarithms}
The resummation of infrared logarithms with next-to-leading
accuracy requires the knowledge of the coefficients of double and
single logarithms at one loop and the coefficient of leading
logarithms at two loops.
\\
At one loop the distribution takes the look
\begin{eqnarray}\label{1loop}
D(x)&=&1-\alpha_S A_1\log^2 x \ +\alpha_S B_1\log x \ +\nonumber\\
&+&\frac{\alpha_S C_F}{\pi}k_1+\frac{\alpha_S C_F}{\pi} r_1(x)
\end{eqnarray}
The coefficients $A_1$ and $B_1$ can be calculated by general
properties of QCD, applying the perturbative evolution for the
light quark, described by the Altarelli-Parisi kernel, and the
eikonal approximation for the massive quark \cite{eikonal}.
\\
The coefficients turns out to be \cite{noi}
\begin{equation}
A_1= {C_F \over \pi} \ \ \ , \ \ \ B_1=-{5\over 4}{C_F\over \pi}
\end{equation}
Let us notice that the coefficient $B_1$ is larger than in
processes involving light quarks only, due to gluon emissions from
the heavy quark and this produces an enhancement of the single
logarithmic term. The two loop coefficient $A_2$ is known from
literature \cite{A2}
\begin{equation}
A_2={C_F\over 2\pi^2} [ C_A ({67\over18} - {\pi^2\over6}) -
{5\over9} \ n_f]
\end{equation}
The resummation of transverse momentum is performed in the space
of impact parameter $b$
\begin{equation}
\tilde{D}(b) = \int_{-\infty}^{+\infty} d\vec{p_\perp}
e^{i\vec{p_\perp}\cdot\vec{b}} \ {1 \over \Gamma_0} \ {d\Gamma
\over d\vec{p_\perp}} \ (\vec{p_\perp})
\end{equation}
which allow the factorization of the phase space and of the
kinematical constraint.
\\
The result for the functions $g_1$ and $g_2$ appearing in
(\ref{g}) is
\begin{eqnarray}
g_{1}(\omega ) &=&{\frac{A_{1}}{2\beta _{0}}}\frac{1}{\omega
}\left[ \log
(1-\omega )+\omega \right]  \\
g_{2}(\omega ) &=&-{\frac{A_{2}}{2\beta _{0}^{2}}}\left[ {\frac{\omega }{%
1-\omega }+}\log (1-\omega )\right] + \nonumber\\ &+&{\frac{A_{1}\beta _{1}}{2\beta _{0}^{3}}%
}[ {\frac{\log (1-\omega )}{1-\omega }+\frac{\omega }{1-\omega
}}+\nonumber \\ &+& {\frac{1}{2}}\log ^{2}(1-\omega )]
+{\frac{B_{1}}{\beta _{0}}}\log (1-\omega )\qquad
\end{eqnarray}
being $\omega = \beta_0 \alpha_S \log ({Q^2 b^2 \over b_0^2})$ and
$b_0=2e^{-2\gamma_E}$.
\\
Let us notice that the resummed distribution diverges for
$\omega\rightarrow 1$ and this signals the appearance of non
perturbative effects, related to the Landau pole of the theory.
They appear when the distribution approaches the limit of the
phase space in the infrared region, that is, roughly speaking, for
$p_\perp \sim \Lambda$.
\\
The resummed distribution is more reliable than the fixed order
calculation near the border of phase space, nevertheless it breaks
down for small value of the transverse momentum.
\\
This divergence should be factorized in a non perturbative
structure function: for the photon spectrum an approach based on
an effective theory has been used, by introducing a \it shape
function \rm \cite{shape}, but in this case we conclude in
\cite{noi} that this approach is not viable and one should try
another way to include non perturbative effects.

\subsection{Fixed order calculation}
A complete resummed distribution with next-to-leading accuracy
requires the calculation of the constant $k_1$ in (\ref{1loop}).
The first term in the expansion of the remainder function,
$r_1(x)$, should be relevant too, due to the low value of the hard
scale. They can be extracted by an explicit calculation of real
and virtual diagrams at one loop.
\\
The process $b\rightarrow s\gamma$ is loop mediated in the
Standard Model: an effective hamiltonian has been developed to
integrate out the heavy particles in the loop \cite{born}
\begin{equation}\label{hamiltoniana}
{\cal H}_{eff} = {G_F \over \sqrt{2}} \ V^*_{ts}V_{tb} \
\sum_{j=1}^8 C_j(\mu) \ \hat{\cal O}_j(\mu)
\end{equation}
where $C_j$ are Wilson coefficients and $\hat{\cal O}_j(\mu)$ are
effective operators ($i=1,\dots,8$)\cite{misiak}.
\\
$\hat{\cal O}_7$ is the most relevant operator because it is
affected by a logarithmic enhancement in the infrared region,
while the other operators contribute to constants \cite{pott}.
\\
Real diagrams are calculated in dimensional re\-gu\-la\-rization
in $n=4+\epsilon$ dimensions.
\begin{eqnarray}\label{integrazione}
D_R(x)= \int_0^1 d\Phi \ \mathcal{M}_{4+\epsilon}(\omega,t) \
\theta[x - \omega^2t(1-t)]&&
\end{eqnarray}
being $\Phi$ the 3-body phase space, $\omega$ the gluon energy
fraction and $t=(1-\cos \theta) /2$, where $\theta$ is the angle
between the gluon and the direction $-\hat{z}$.
$\mathcal{M}_{4+\epsilon}(\omega,t)$ is the matrix element with
the insertion of $\hat{\cal O}_7$ in the vertex, in dimensional
re\-gu\-la\-rization.
\\
The integration may be analytically performed by introducing
harmonic polylogarithms as in \cite{rv}, satisfying the properties
\begin{eqnarray}
H(a;y)&=&\int_0^y dy^\prime \ g(a;y^\prime) \ \ \  a\not= 0\nonumber\\
H[0,y]&=& \log y\nonumber\\
H(\vec{m}_w;y) &=& \int_0^y dy^\prime \ g(a;y^\prime) \ H(\vec{m}_{w-1};y^\prime)\nonumber\\
{d\over dy} H(\vec{m}_w;y)&=&g(a;y)H(\vec{m}_{w-1};y)
\end{eqnarray}
where the basis of functions $g(a;y)$ is
\begin{eqnarray}\label{base}
g[0;y]&\equiv& {1 \over y} \nonumber\\ g[-1;y]&\equiv&
{1 \over y+1} \nonumber \\ g[-2;y]&\equiv& {1 \over {\sqrt{y}(1+y)}}\nonumber \\
g[-3;y]&\equiv& -{\sqrt{x} \over
{2({1-\sqrt{x}\sqrt{y})}\sqrt{y}}}
\end{eqnarray}
Details of the calculation and the explicit result for the first
term in the expansion of the remainder function are shown in
\cite{noi2}.
\\
Virtual diagrams are calculated with ordinary techniques: they
cancel the infrared poles a\-ri\-sing from real diagrams and don't
depend on the kinematics, giving contribution to the coefficient
function only. In particular the vertex correction is calculated
by reducing the diagram to simpler topologies and using
integration by parts identities: details of the calculation are
discussed in \cite{noi2}.
\\
Summing real and virtual diagrams the first term of the
perturbative expansion of the coefficient function for the
operator $\hat{\mathcal{O}}_7$ turns out to be
\begin{equation}
k_1={C_F \over \pi} \ (- \ {11 \over 4} \ - {\pi^2 \over 12}\ +\
\log{m_b\over \mu})
\end{equation}
An improved result for the coefficient function, taking into
account the other operators of the basis (\ref{hamiltoniana}), can
be found in \cite{noi2}.

\section{Conclusions and Outlook}
$b\rightarrow s\gamma$, though a rare decay, seems to be a very
clean process to study hadronic physics in $b$ decays and to apply
resummation techniques to check whether they are reliable in $b$
physics or not. A very accurate calculation of all the ingredients
involved in the resummation "recipe" is needed to this aim.
Otherwise, it would be impossible to conclude whether
discrepancies from experimental data may be attributed to this
approach or to other sources of uncertainty typical of $b$
physics, such as an imperfect knowledge of CKM matrix elements or
the $b$ quark mass.
\\
A more interesting application should concern the transition
$b\rightarrow c$, which is phe\-no\-me\-no\-lo\-gi\-cal\-ly more
relevant, but whose calculation is complicated by the presence of
the $c$ quark mass.
\\
From this point of view the study of $b\rightarrow s\gamma$ may be
seen as a first step to apply resummation techniques to $b$
physics, before going to more complicated but more important
processes.
\\
\\
The contents of this paper arise from my work with Prof.
L.Trentadue and Dr. U.Aglietti: I wish to thank Prof. Trentadue
for his collaboration and Dr. Aglietti for his essential help.

\end{document}